\RequirePackage{lineno}
\documentclass[%
 reprint,
twocolumn,
showpacs,preprintnumbers,showkeys,
 amsmath,amssymb,
 aps,
prc,
floatfix,
]{revtex4-2}

\usepackage{graphicx}% Include figure files
\usepackage{dcolumn}% Align table columns on decimal point
\usepackage{bm}% bold math
\usepackage{amssymb}
\usepackage{float}
\usepackage[caption = false]{subfig}
\usepackage{hyperref}% add hypertext capabiliti
\begin{document}

%\preprint{APS/123-QED}

\title{Ridge from jet-medium interaction in p-Pb collisions at $\sqrt{s_{NN} }$ = 5.02 TeV}
\author{Debojit Sarkar}
\email{debojit03564@gmail.com}
\author{Subhasis Chattopadhyay}

\affiliation{Variable Energy Cyclotron Centre, HBNI, 1/AF-Bidhannagar, Kolkata-700064, India}
\date{\today}

%\corauth[cor]{Corresponding author.}
%\ead{sub.chattopadhyay@gmail.com}
%\address[label1]{Variable Energy Cyclotron Centre, 1/AF-Bidhannagar, Kolkata-700064, India}

\begin{abstract}
In this paper we report the effect of the jet-medium interplay as implemented in  EPOS 3 on the ridge like structure observed in high multiplicity p-Pb collisions at $\sqrt{s_{NN}} = $ 5.02 TeV. EPOS 3 takes into account hydrodynamically expanding bulk matter, jets and the jet-medium interaction. The basis of this model is multiple scatterings where each scattering finally produces flux tube / string. In the higher multiplicity event classes where the flux tube/string density is higher, there is a finite probability that the strings will pick up quarks and antiquarks (or diquarks) from the bulk (core) for flux tube breaking to produce jet hadrons (corona) instead of producing them via usual Schwinger mechanism. This will eventually create a correlation between core and corona and also influence the corona-corona correlation as the corona particles containing quarks and antiquarks (or diquarks)  from the bulk also carry the fluid information. We report the relative contributions of the core-core, core-corona, corona-core and corona-corona correlations towards the ridge in the high and low multiplicity p-Pb collisions at $\sqrt{s_{NN}} = $ 5.02 TeV using the data generated by EPOS 3. The multiplicity evolution of the ridges in all the cases is also reported.
\end{abstract}

%\pacs{24.30.Cz, 29.40.Mc, 24.60.Dr.}
%\begin{keyword}

\keywords{EPOS 3; core-corona separation; fluid-jet interaction; hydrodynamics, ridge } 

%\end{keyword}
%\end{frontmatter}

\maketitle

\section{Introduction}

Two-particle angular correlation measurements in p-Pb \cite {alice_pPb_double_ridge}, pp \cite {CMS_pp_Ridge} and d-Au \cite {d-Au ridge paper} collisions have revealed existence of
azimuthal correlations extended to large pseudorapidity separation $|\Delta\eta|$ popularly known as ''ridge''. In heavy-ion collisions, such long-range structures along $|\Delta\eta|$
have been attributed to the collective emission of  particles representing the initial anisotropy in the over-lap geometry of the colliding nuclei.
Several other observations, like,
 mass-ordering of the elliptic flow coefficient ($v_{2}$) of identified particles  \cite {pPb_mass_ordering}, quark-scaling of $v_{2}$ \cite {CMSv2paper} and baryon-to-meson enhancement at intermediate $p_{T}$  \cite {p_pi_enhancement_pPb} suggest
similarity between the systems formed in small and heavy-ion collisions.
It has been argued that angular-correlations in small systems are dominated by jet-like processes. However, the emergence of the near side ridge in high-multiplicity event classes of small collision systems (pp and p-Pb) \cite {CMS_pp_Ridge}  \cite {pPb_mass_ordering} still lacks un-ambiguous understanding. Several theoretical propositions have been made based on correlated emission from
 glasma flux tubes(CGC approach)  \cite {CGC_pPb}, collective flow due to hydro-dynamical effects \cite {epos_massordering_flow_pPb} \cite {epos_ridgein_pp} or incohrent parton scatterings \cite {AMPT_pp_pPb_ridge} \cite {AMPT_pPb_flow}, but no general agreement could be reached. EPOS 3 creates significant ridge like structure in the higher multiplicity classes of pp \cite {epos_ridgein_pp} and p-Pb \cite {epos_massordering_flow_pPb} collisions at LHC energy and it has been discussed in terms of hydrodynamical evolution of the medium as the ridge structure vanishes in case the hydro evolution is switched off \cite {epos_ridgein_pp}. This indicates that the hydrodynamical evolution as implemented in EPOS 3 is the key ingredient to create the ridge structure in high multiplicity classes of small collision systems. Our work shows that in the context of EPOS 3, ridge in the higher multiplicity classes of p-Pb collisions has non zero contribution even from the fluid-jet interaction \cite {epos_PbPb_lambdakshort_enhance}. \\

EPOS 3 is a 3+1D event by event hydro model based on flux tube initial conditions \cite {epos_radialflow_spectra_pPb}  \cite {epos_model_descrip}. The basis of this model is multiple scatterings where each scattering consists of a hard elementary scattering plus initial state radiation - commonly referred as a parton ladder or pomeron. In this formalism each ladder may be considered as a longitudinal colour field which can be treated as a relativistic string. After multiple scatterings the final state partonic system consists of mainly longitudinal colour flux tubes carrying transverse momentum of the hard scattered partons in the transverse direction- known as kinks in the string language. Depending on the energy of the string segments and local string density, the high density areas form the "core"(containing string segments more than a critical value per unit area in  given transverse slices) and the low density area form the "corona" \cite {epos_core_corona_sep}. The strings in the core thermalize and then undergo hydrodynamical expansion and finally hadronize to form the bulk part of the system. In the low density region, the flux tubes (strings) eventually expand and finally break via the production of quark-antiquark or diquark-antidiquark pairs following Schwinger mechanism responsible for production of jet hadrons (corona) in EPOS.  But in the high multiplicity event  classes where the density of the string segments is higher because of many elementary scatterings, the strings  cannot decay indepenedently following Schwinger mechanism only. After initial scatterings, the produced flux tubes / strings initially form a "matter" which eventually constitute both bulk  and jets based on the energy loss by them. As mentioned in \cite {epos_PbPb_lambdakshort_enhance} three possibilities can occur:

a) String segments without sufficient energy to escape the matter will evolve hydrodynamically and finally hadronize to produce the bulk (core).\\
\\
b) High energetic string segments will escape the matter and hadronize following Schwinger mechanism producing jet hadrons or corona.\\
\\
c) Some string segments are produced inside the matter or at the surface but have enough energy to escape. These segments may pick up quark, antiquark, diquark or antidiquark needed for the flux tube breaking from the fluid (bulk) with properties (momentum, flavor) determined by the fluid \cite {epos_PbPb_lambdakshort_enhance} rather than the Schwinger mechanism. The produced jet hadrons are composed of a high $p_{T}$ string segment originating from the initial hard process and di(quarks) from the fluid carrying fluid properties including transverse fluid velocity. As a result, these jet hadrons appear at relatively higher $p_{T}$ than the jet hadrons produced via usual Schwinger mechanism only and also provide information about the fluid \cite {epos_PbPb_lambdakshort_enhance}.
As baryons carry higher number of quarks compared to mesons, they will have larger momentum push from the fluid-jet interaction contributing to the inclusive baryon to meson enhancement at intermediate $p_{T}$ as observed in Pb-Pb collisions at $\sqrt{s_{NN}} = $ 2.76 TeV \cite {epos_PbPb_lambdakshort_enhance}. It has also been shown in \cite {epos_PbPb_lambdakshort_enhance} that this effect is more pronounced in the higher multiplicity classes of Pb-Pb collisions at $\sqrt{s_{NN}} = $ 2.76 TeV where the transverse size of the system is higher increasing the probability of fluid-jet interaction.\\ 
A similar baryon to meson enhancement was also observed in p-Pb collisions at $\sqrt{s_{NN}} = $ 5.02 TeV and EPOS 3 can qualitatively generate the pattern \cite {epos_radialflow_spectra_pPb}.
The jet hadrons  carrying fluid properties (produced via formalism (c)) are expected to be correlated with the bulk part of the system. Therefore except from core-core , core-corona and corona-core correlations are expeceted. Also, it would be interesting to look at the corona-corona correlations at higher multiplicity event class where a larger fraction of corona particles are expected to be formed inside the matter \cite {epos_PbPb_lambdakshort_enhance} and carry fluid properties compared to the lower multiplicity event classes.\\ 
In this paper we have adopted the two-particle correlation technique to investigate the multiplicity evolution of the near-side ridge with triggers (2.0  $<p_{T}< $4.0 GeV/$\it{c}$) and associateds (1.0  $<p_{T}< $4.0 GeV/$\it{c}$) from intermediate $p_{T}$ in p-Pb collisions at 5.02 TeV. $p_{T}$ ranges of trigger and associated particles are chosen in the region where the ridge structure has been prominently observed in the experimental results \cite {alice_pPb_double_ridge}. An inclusive baryon to meson enhancement has also been observed in this $p_{T}$ range in p-Pb collisions at 5.02 TeV \cite {p_pi_enhancement_pPb}. In this analysis, the ridge is defined by $|\Delta\eta|$ separation ($|\Delta\eta|$ $\geq$ 1.2) in the near side 
 ($|\Delta\phi|$  $< \pi/2$) of the correlation function  as it is done in  \cite {MPI_pPb}. The multiplicity evolution of the ridges in core-core, core-corona, corona-core and corona-corona correlations has been studied. The relative contribution of the core-core (pure hydrodynamical origin), core-corona and corona-core (fluid-jet interaction) and corona-corona correlations towards the "total" ridge in the high multiplicity event class of p-Pb collisions has also been investigated.

\begin{figure}[htb!]
%\begin{center}
a)

\includegraphics[scale=0.32]{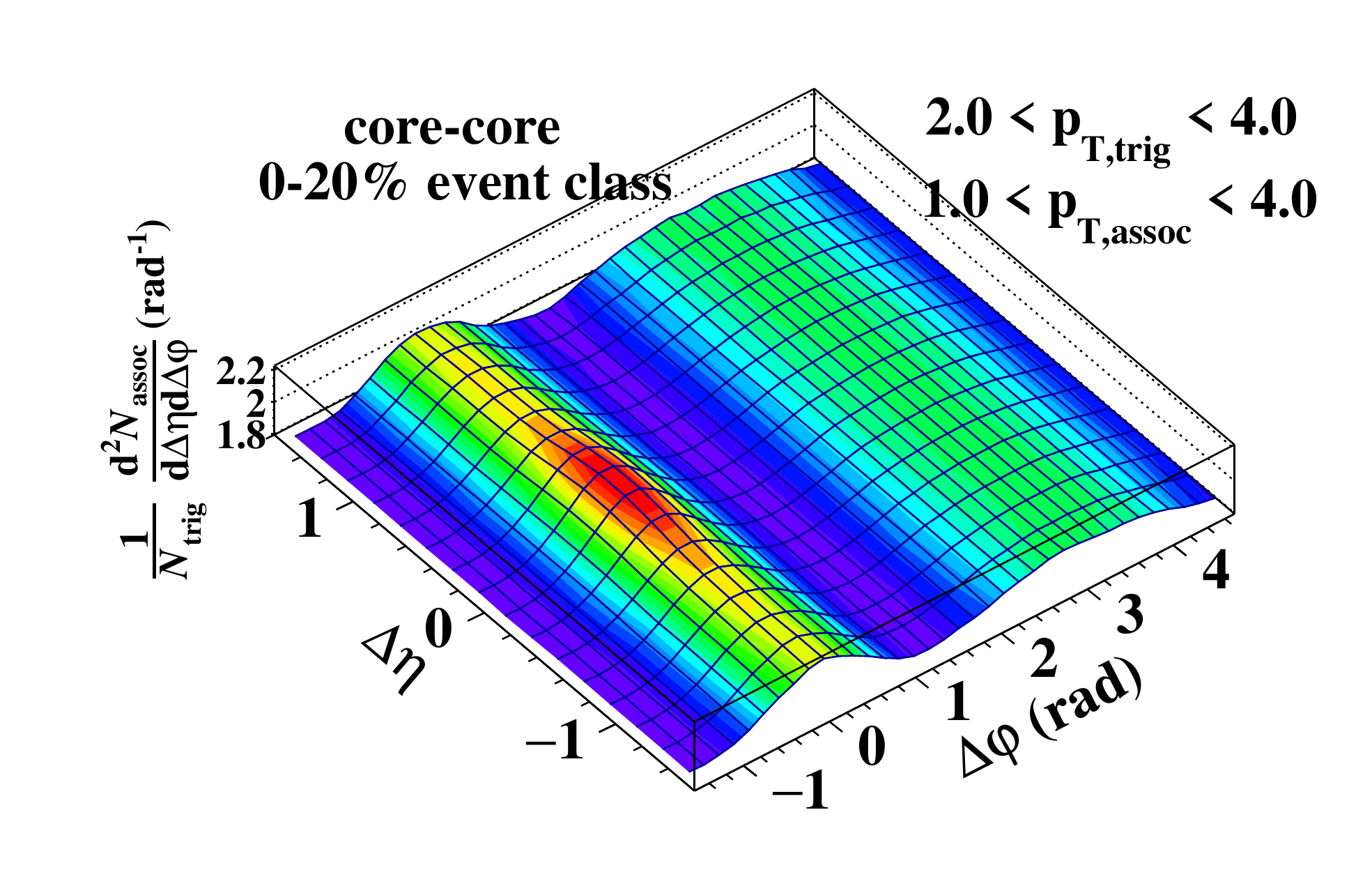}

b)

\includegraphics[scale=0.33]{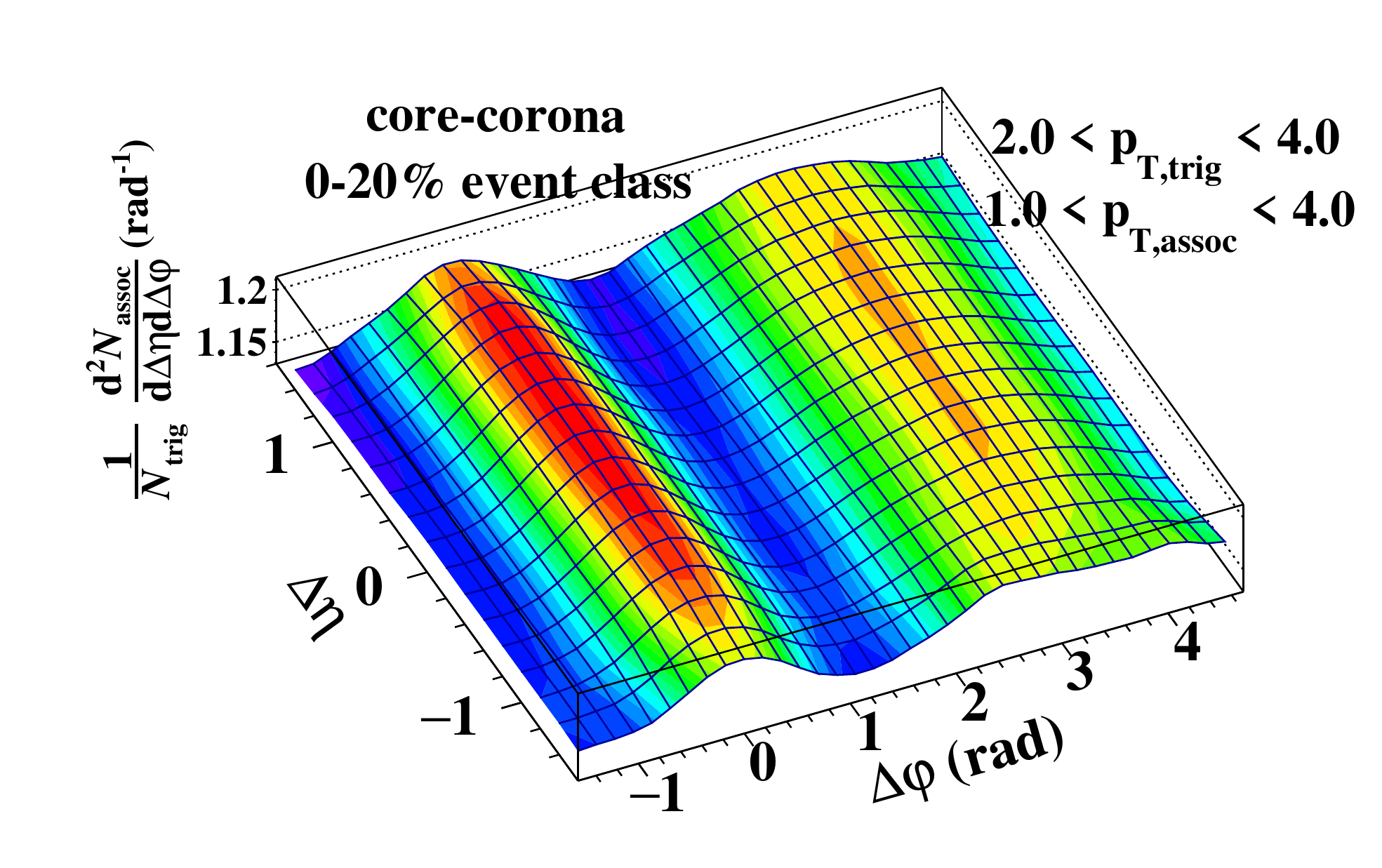}

c)

\includegraphics[scale=0.32]{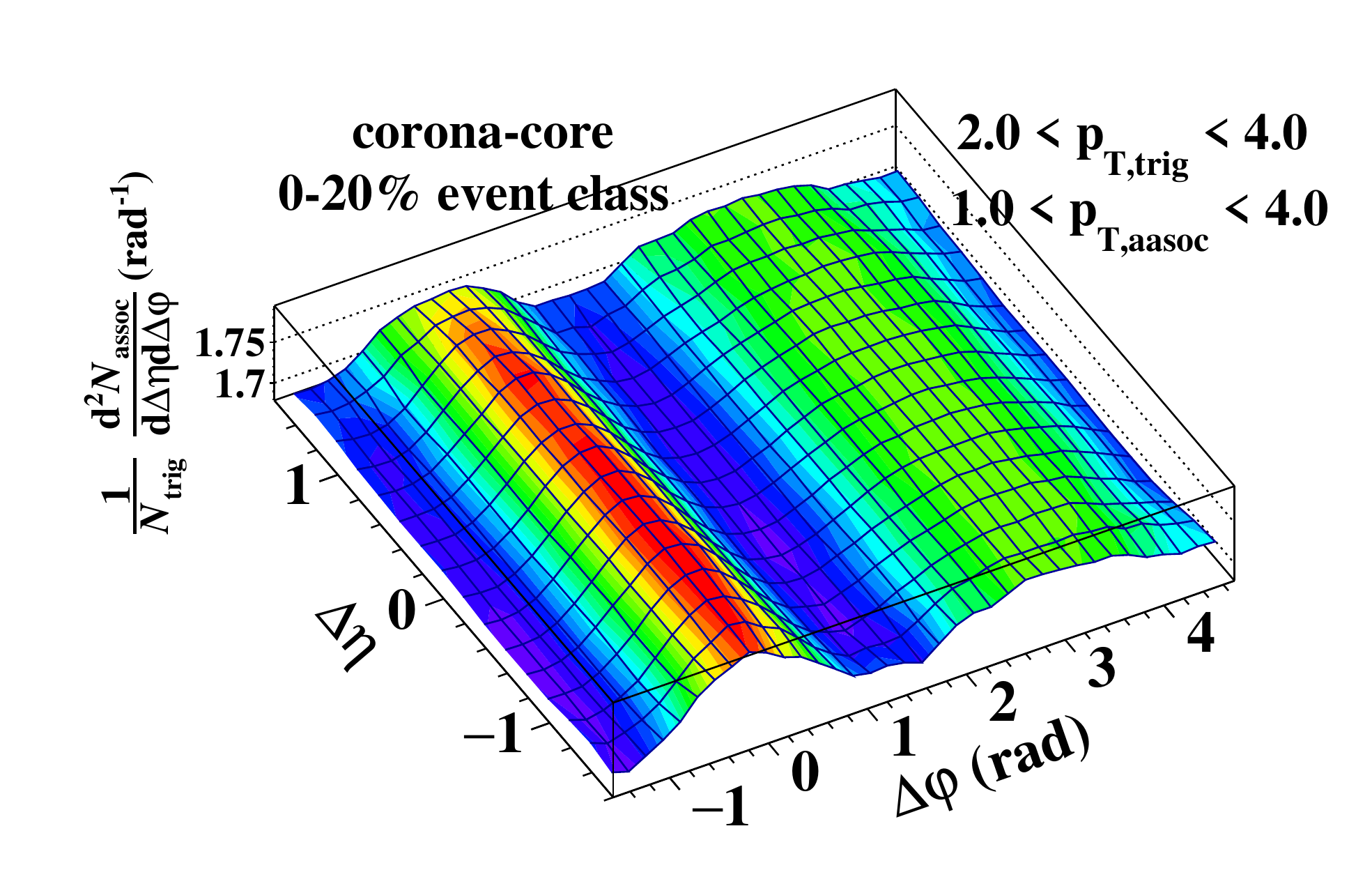}

d)

\includegraphics[scale=0.32]{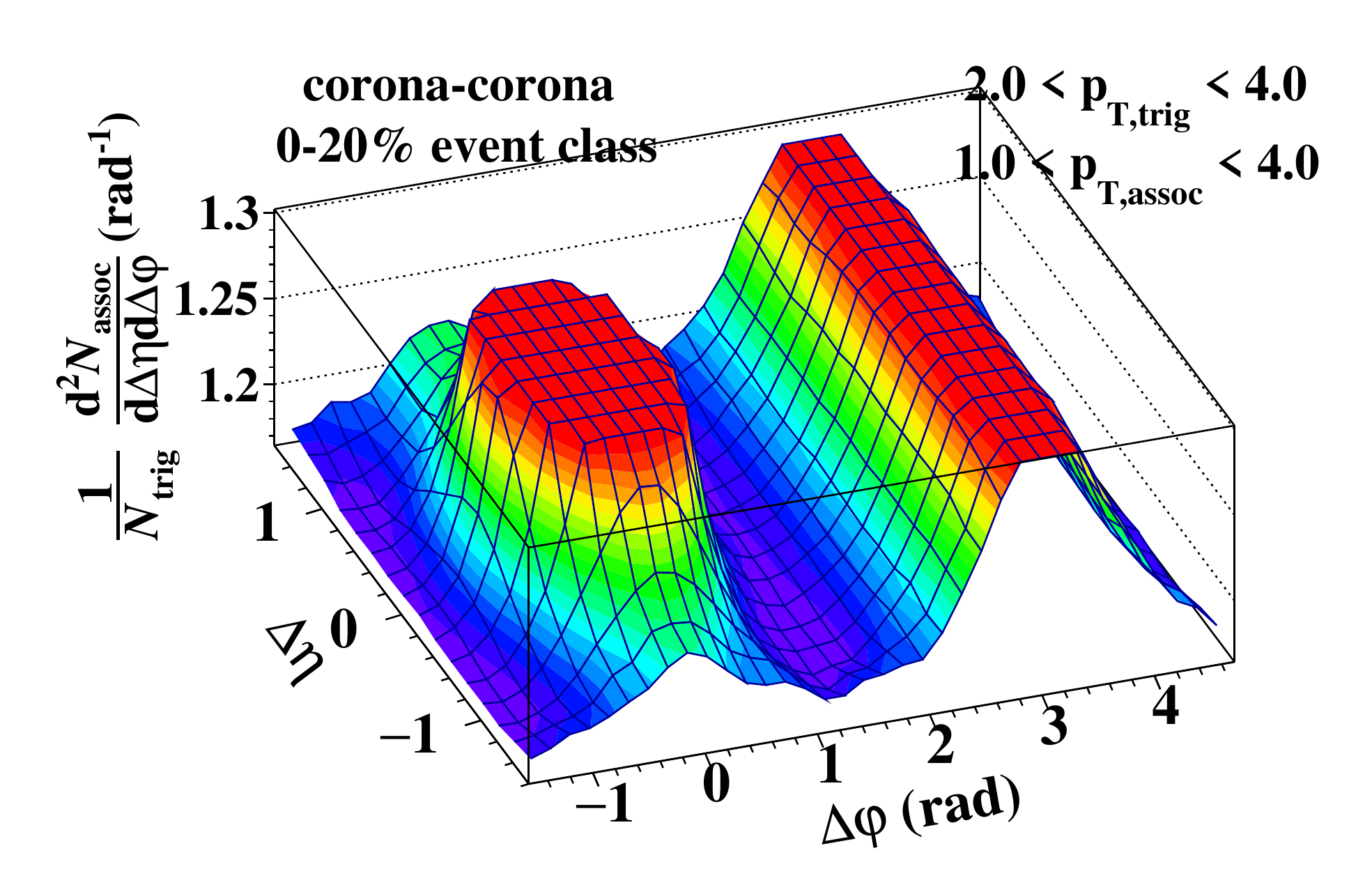}

\caption{[Color online] Two Particle $\Delta\eta$-$\Delta\phi$ correlation function in the 0-20 $\%$ event class 
of p-Pb collisions at $\sqrt{s_{NN} }$ = 5.02 TeV from EPOS 3 for  a) core-core  b) core-corona  c) corona-core  d) corona-corona correlations.}
%\end{center}
\label{partondist_XY}
\end{figure}

\section{The EPOS 3 Model}

EPOS 3 basically contains a hydrodynamical approach based on flux tube initial conditions \cite {epos_radialflow_spectra_pPb}  \cite {epos_model_descrip} \cite {parton_gribov_Regge_Th}. This is a parton based  model where the partons initially undergo multiple scatterings. Each scattering is composed of  hard elementary scattering with initial and final state linear parton emission- forming parton ladder or "pomeron". Each parton ladder has it's own saturation scale $Q_{s}^{2}$ depending on the number of connected participants and it's center of mass energy and this will separate the soft processes from hard/p-QCD ones. This formalism is referred as "Parton based Gribov Regge Theory" and explained 

%[htb!]
\begin{figure}
%\begin{center}
a)

\includegraphics[scale=0.30]{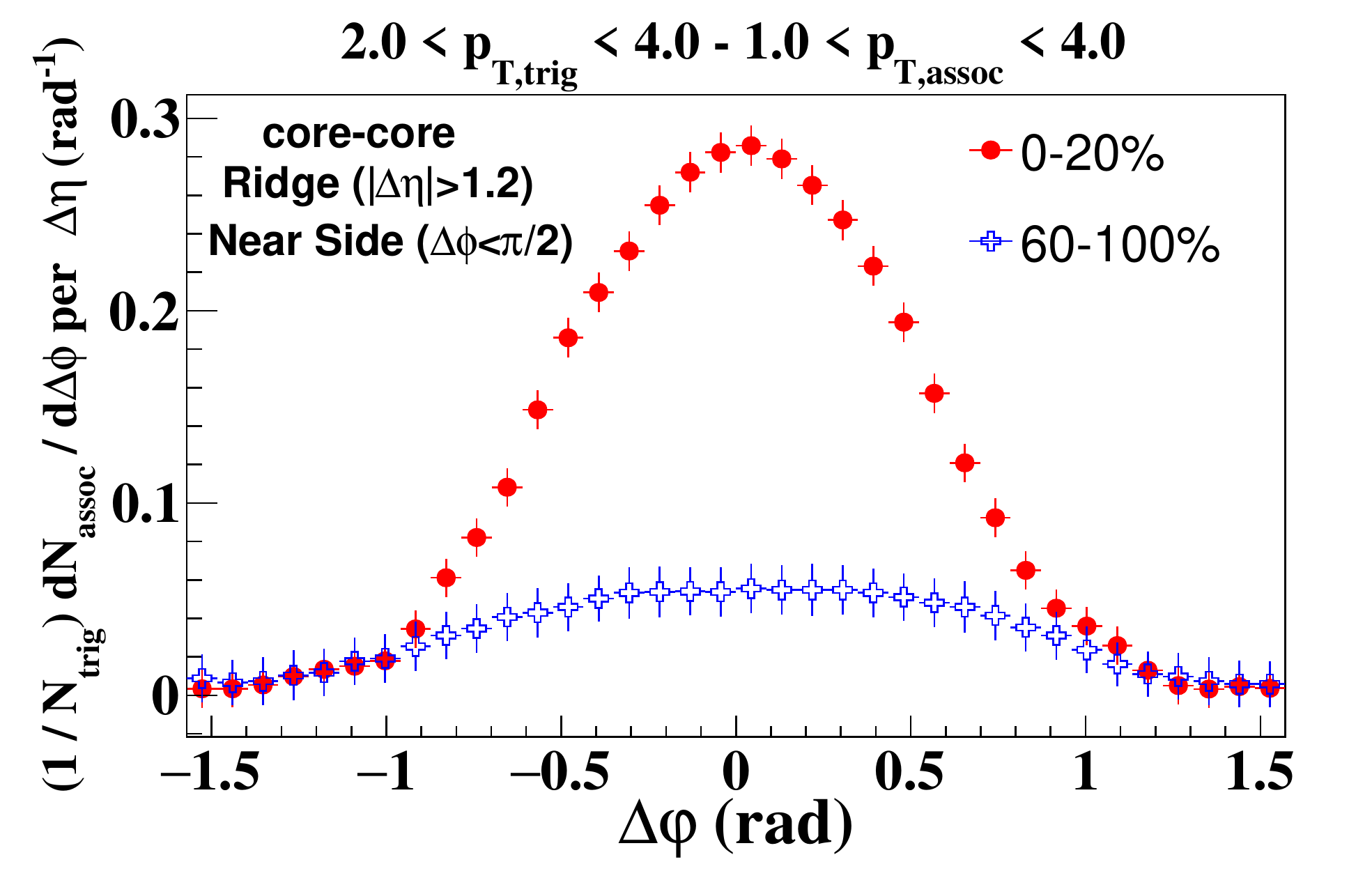}

b)

\includegraphics[scale=0.31]{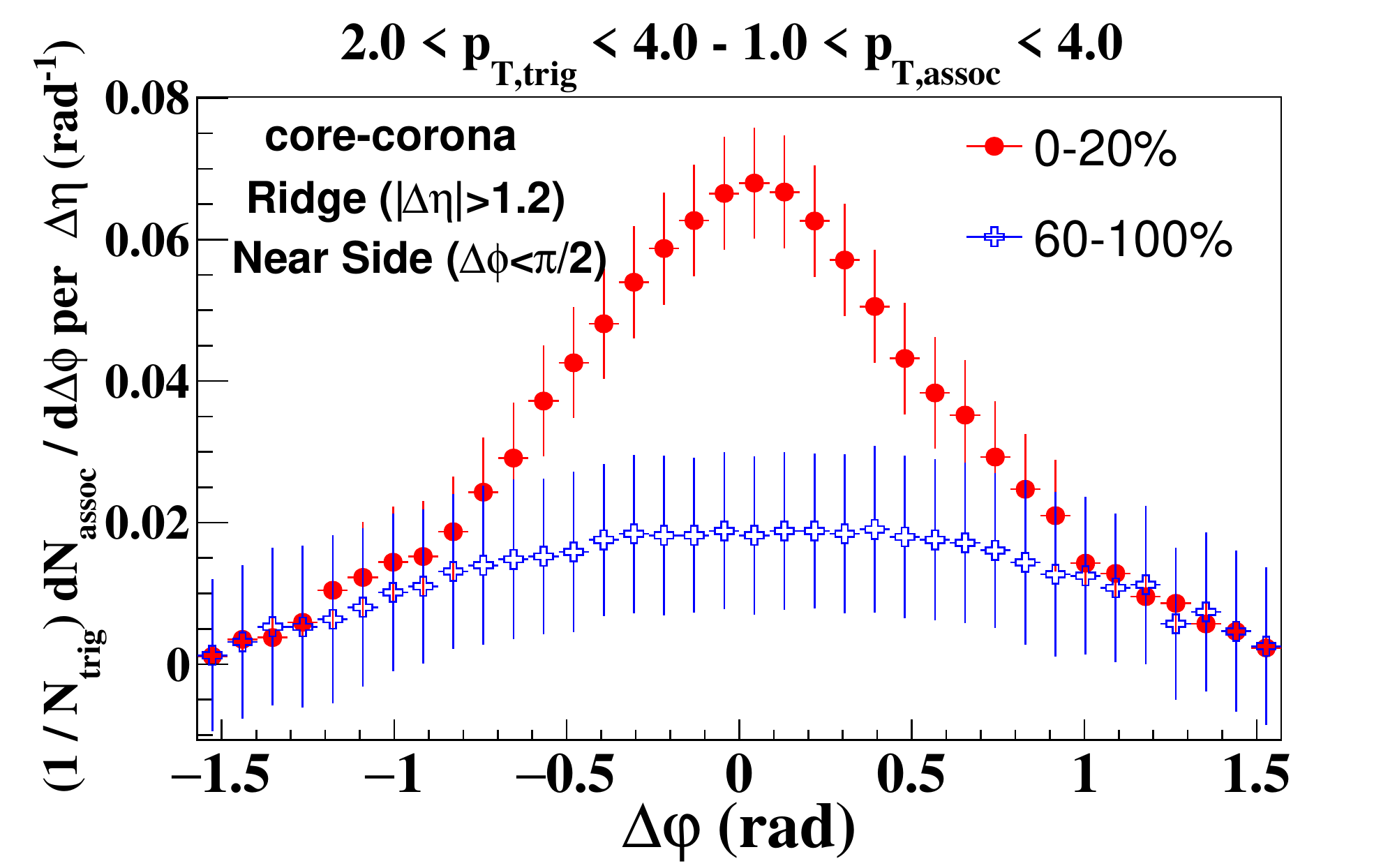}

c)

\includegraphics[scale=0.31]{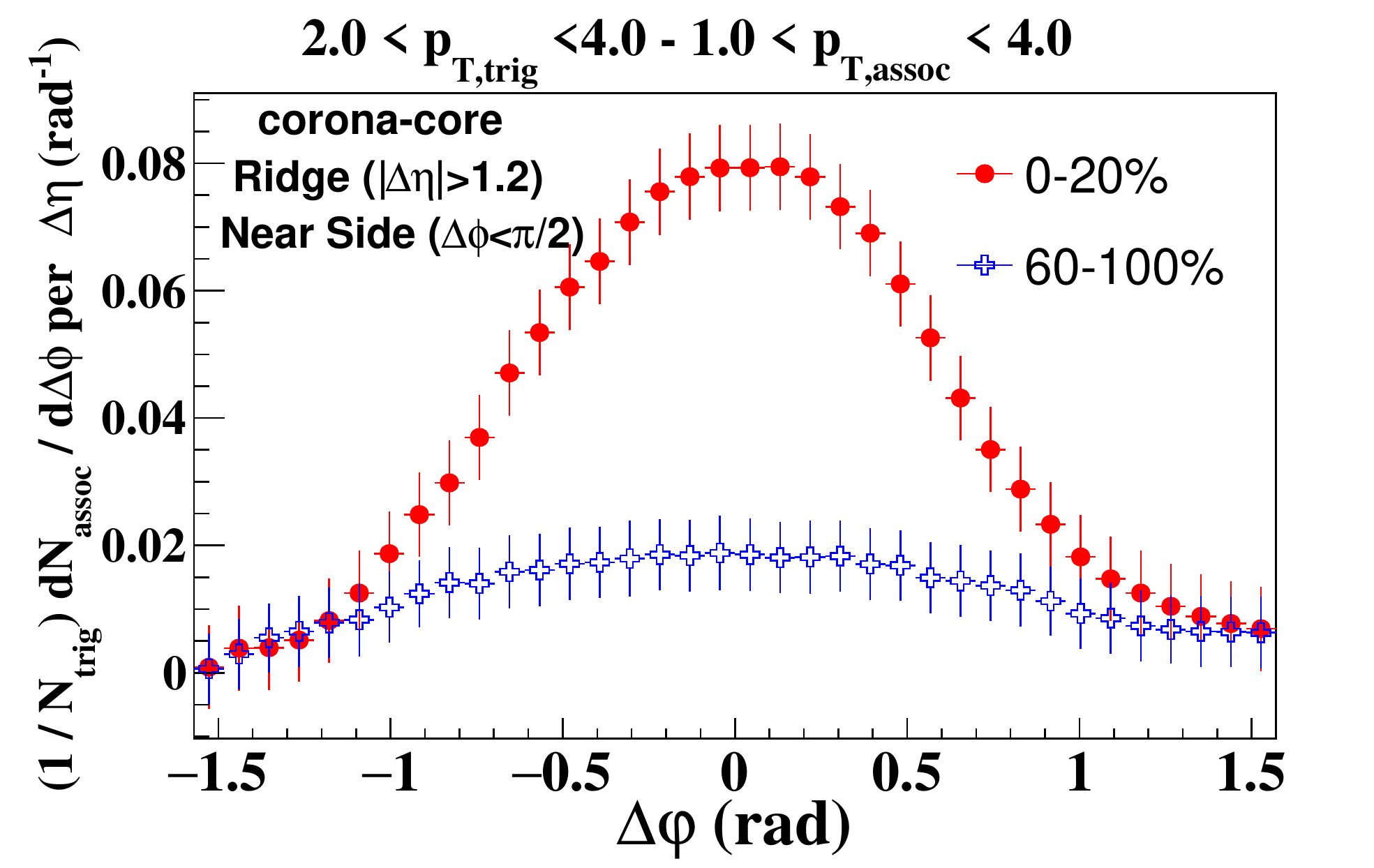}

d)

\includegraphics[scale=0.31]{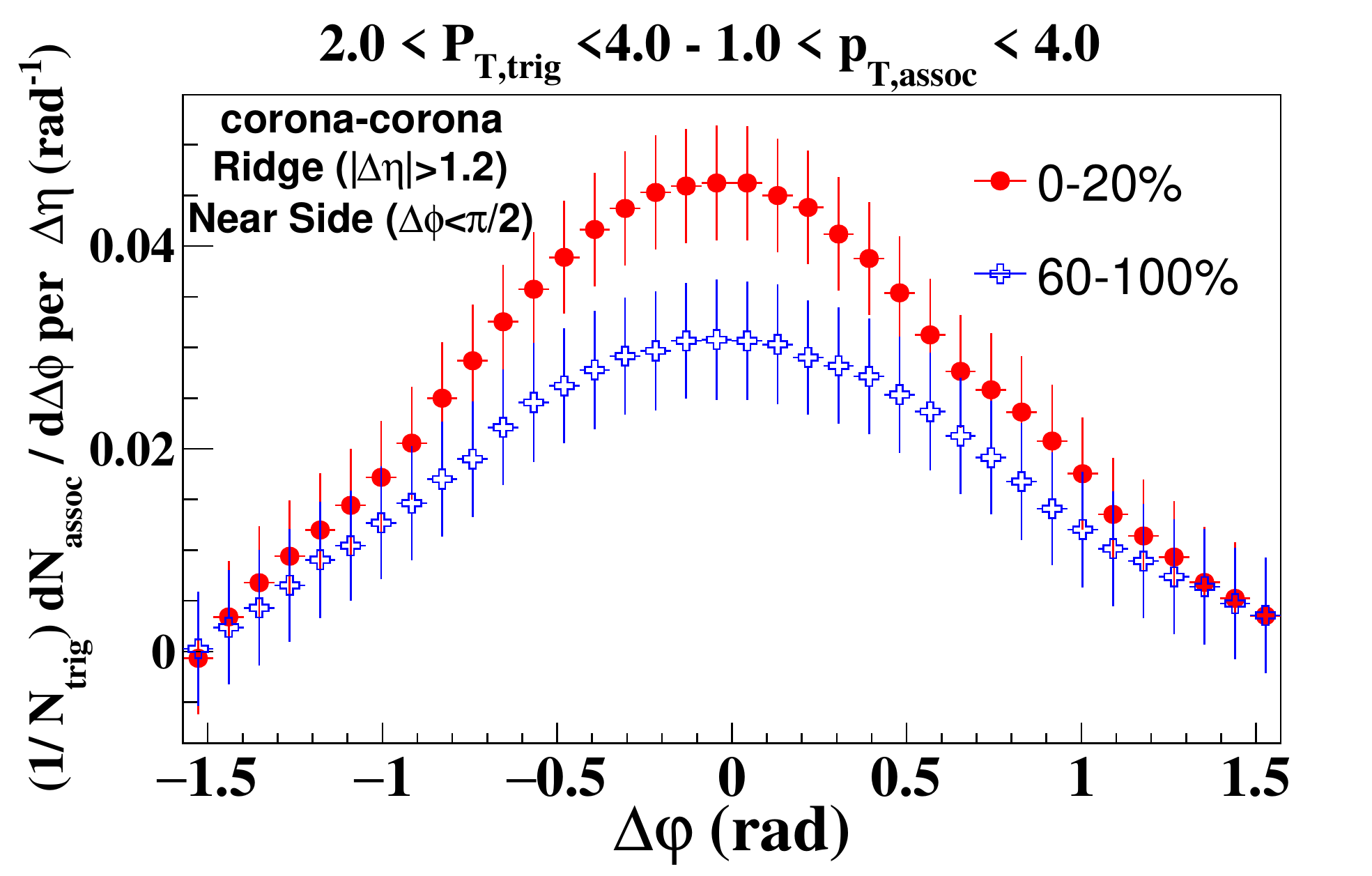}

\caption{[Color online] Multiplicity dependence of the near-side ridge (pedestal subtracted) obtained from a) core-core  b) core-corona  c) corona-core  d) corona-corona correlations. In each case ridges from 0-20\% and 60-100\% event classes have been compared.}
%\end{center}
\label{partondist_XY}
\end{figure}
in detail in \cite {epos_massordering_flow_pPb}  \cite {parton_gribov_Regge_Th}. In this formalism each ladder may be considered as a longitudinal colour field which can be treated as a relativistic string. After multiple scatterings the final state partonic system consists of mainly longitudinal colour flux tubes carrying transverse momentum of the hard scattered partons in transverse direction. Depending on the energy of the string segments and local string density - the high density areas form the "core" and the low density area form the "corona" \cite {epos_core_corona_sep}. The core part basically forms the bulk matter. The strings in the core thermalize and then undergo hydrodynamical expansion and finally hadronize. The strings in the corona hadronize by Schwinger's mechanism and basically constitutes the jet part of the system. EPOS 3 has reasonably reproduced the double ridge and mass ordering of $v_{2}$ of identified particles in high multiplicity p-Pb collisions \cite {epos_massordering_flow_pPb} as observed by the ALICE collaboration at CERN \cite {alice_pPb_double_ridge} \cite {pPb_mass_ordering}. 
In section VIII of \cite {epos_radialflow_spectra_pPb}, the multiplicity dependence of the identified particle spectra (as measured by ALICE in p-Pb collisions at 5.02 TeV) has been compared with different event generators (e.g. QGSJETII, AMPT and EPOS 3). The radial flow generated in a hydro model like EPOS 3 pushes heavier particles more compared to the lighter ones from lower to higher $p_{T}$ values describing the spectra at intermediate $p_{T}$ in a better way \cite {epos_radialflow_spectra_pPb}. Furthermore the multiplicity dependence of the particle ratios measured by ALICE has been compared with the output of different event generators mentioned above \cite {epos_radialflow_spectra_pPb}. The trend of the baryon to meson enhancement at intermediate $p_{T}$ has been observed in EPOS 3 \cite {epos_radialflow_spectra_pPb}. In addition to the radial flow, the fluid-jet interaction as implemented in EPOS 3 also contributes to the baryon to meson enhancement at intermediate $p_{T}$ \cite {epos_PbPb_lambdakshort_enhance} and has already been discussed in the previous section.\\
The effect of the fluid-jet interaction on the ridge structure observed in high multiplicity p-Pb collisions at 5.02 TeV will be discussed in this paper.

\section{Analysis Method}

Two particle correlation is a widely used technique in high-energy physics for extracting the properties of the system 
produced in collisions at ultra relativistic energies. In such studies, the correlation function is obtained 
in terms of the difference in the azimuthal angle and pseudo-rapidity of two sets of particles
classified as {\it trigger} and {\it associated}. The p${_T}$ range of trigger and 
associated particles are 2.0 $<p{_T}<$ 4.0 GeV/$\it{c}$ and 1.0 $<p{_T}<$ 4.0 GeV/$\it{c}$ respectively and the correlation function has been constructed with a $p_{T}$ ordering ($p_{T}^{assoc} < p_{T}^{Trigger}$). The pseudo-rapidity of the particles are 
restricted within -0.8$<\eta<$0.8.  A 2D correlation function is obtained as a function of the difference in 
azimuthal angle $\Delta\phi$ = $\bf\phi_{trigger}$ -$\bf\phi_{associated}$ and pseudo-rapidity $\Delta\eta$
= $\bf\eta_{trigger}$ -$\bf\eta_{associated}$. The same event correlation function is defined as 
$\frac{dN_{same}}{N_{trigger}d\Delta\eta d\Delta\phi}$, where $N_{same}$ is the number of particles associated to 
triggers particles $N_{trigger}$. To correct for pair acceptance the same event correlation function is divided by mixed event correlation function  $\alpha$$\frac{dN_{mixed}}{d\Delta\eta d\Delta\phi}$. The mixed event correlation function is constructed by correlating the trigger particles in one event to the associated particles from other events belonging to the same multiplicity event class. The factor $\alpha$ is used to normalize  the mixed event to make it unity for pairs where both particles go into approximately the same direction ($|\Delta\eta|$ $\approx$ 0, $|\Delta\phi|$ $\approx$ 0).

This correlation analysis has been performed by dividing the entire minimum bias events
into different multiplicity classes based on the total amount of charged particles produced (with $p_{T}$ $>$0.05 GeV/c) within 2.8  $<\eta<$5.1. This corresponds to the acceptance range of ALICE VZERO-A detector in the Pb going direction in case of p-Pb collisions and used for multiplicity class determination by the ALICE collaboration \cite {alice_pPb_double_ridge},\cite {pPb_mass_ordering}. But in ALICE, the VZERO detector, an array of 32 scintillator tiles, covering the full azimuth within 2.8  $<\eta<$5.1 (VZERO-A) doesn't give any $p_{T}$ information. Whereas, in this work, only particles with $p_{T}$ $>$0.05 GeV/c are considered even for the multiplicity class determination. Also, in EPOS 3.107 the multiplicity distribution has a tail which is different from data and has been fixed in EPOS 3.117. But we are using EPOS 3.107 and the 0-20\% multiplicity class in this work is not exactly equivalent to the 0-20\% event class in the data.
The multiplicity classes are denoted as 60-100$\%$, 40-60$\%$, 20-40$\%$, 0-20$\%$ from the lowest to the highest multiplicicty.
Two particle correlation functions in case of core-core, core-corona, corona-core and corona-corona correlations in the 0-20\% multiplicity class are shown in Fig 1.

This analysis concentrates only on the near side ($|\Delta\phi|$  $< \pi/2$) of the correlation function.
The particles from jet fragmentation are expected to be confined in a small angular region- so the ridge is estimated from large $|\Delta\eta|$ ($|\Delta\eta|$ $\geq$ 1.2) as it is done in  \cite {MPI_pPb}. The pedestal subtracted ridge structures from the highest (0-20\%) and the lowest (60-100\%) multiplicity classes are compared in Fig 2. The pedestal is determined from the $\Delta\phi$ projection (for $|\Delta\eta|$ $\geq$ 1.2) with the zero yield at minimum (ZYAM) assumption \cite {ZYAM} and estimated as an average of the 8 (out of 36) lowest $\Delta\phi$ points \cite {ZYAM}.

\section{Results and Discussion}

In this analysis the trigger (2.0  $<p{_T}<$ 4.0 GeV/$\it{c}$) and associated (1.0  $<p{_T}<$ 4.0 GeV/$\it{c}$) particles are selected from intermediate $p_{T}$ range where particles from both hard (origin: corona) and soft (origin: core) processes are present \cite {epos_radialflow_spectra_pPb}. In the highest multiplicity event class (0-20\%) where the core contribution and so the fluid-jet interaction are significant, the 2D correlation structure for core-core, core-corona, corona-core and corona-corona correlations are shown in Fig 1. The near side "ridge"(elongated structure along $|\Delta\eta|$) is present in all the cases. In case of core-core correlation (i.e both trigger and associated particles are from core), the ridge is purely from hydrodynamical origin and it increases with multiplicity as shown in Fig 2(a), where the ridge contribution from 60-100\% and 0-20\% event classes are compared.

The ridge structure is also observed in core-corona and corona-core correlations and shown in Fig. 1(b) and Fig 1(c) for 0-20\% event class. The origin of ridge in these two cases is the fluid-jet interaction. 
The jet hadrons (corona) produced inside or at the surface of the bulk via flux tube breaking using partons from the bulk (core) \cite {epos_PbPb_lambdakshort_enhance} carry fluid information and eventually correlated with the bulk part of the system - creating the correlation between core and corona. The probability of fluid-jet interaction increases with increase in multiplicity / system size as more jet hadrons (corona) are expected to be produced using partons from the bulk (core) \cite {epos_PbPb_lambdakshort_enhance}. The multiplicity evolution of the ridge originating from the fluid-jet interaction (core-corona and corona-core correlations) is shown in Fig 2(b) and 2(c) and the ridge contribution increases with increase in multiplicity.

\begin{figure}[htb!]
%\begin{center}
\includegraphics[scale=0.38]{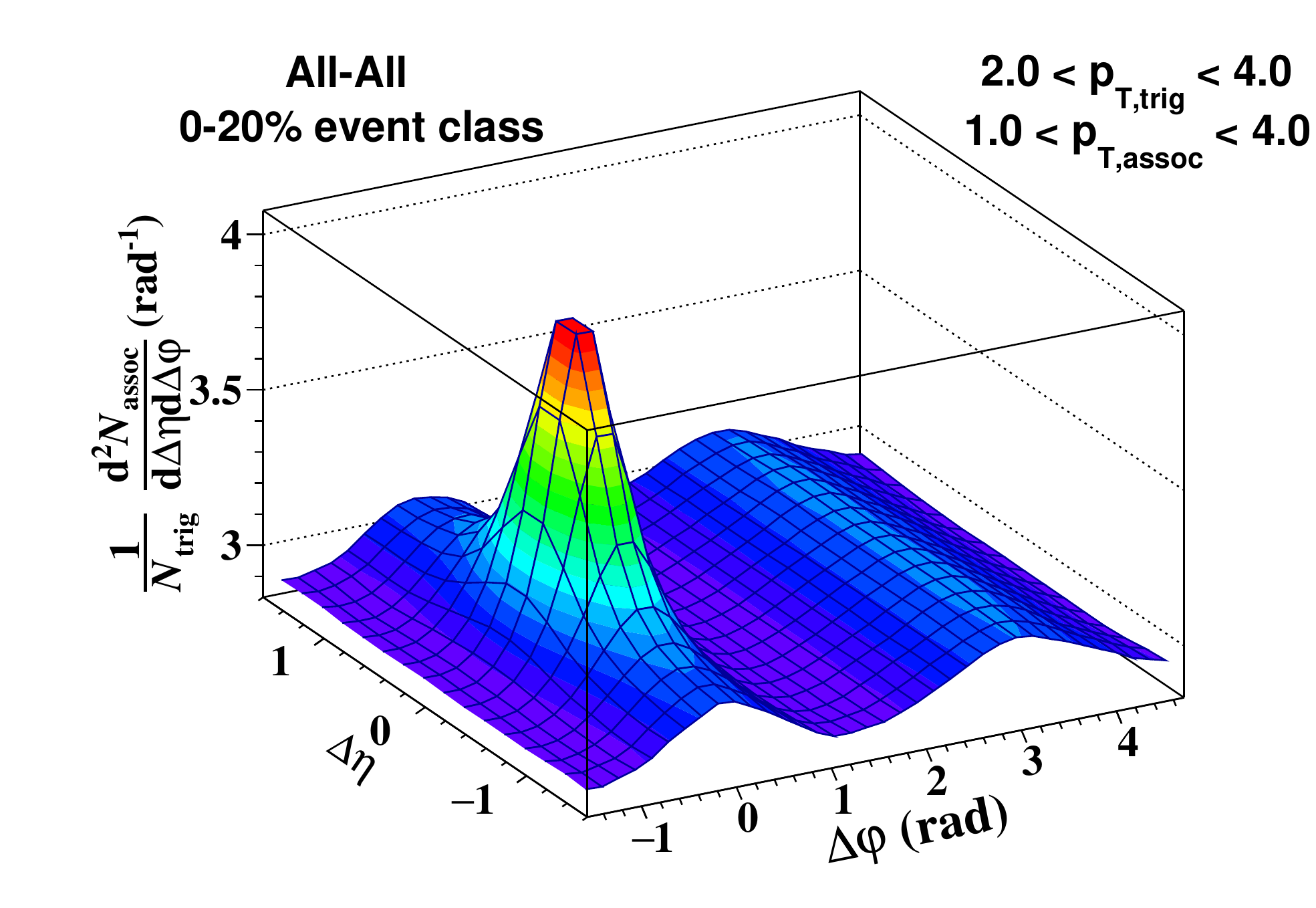}

\caption{[Color online] Two Particle $\Delta\eta$-$\Delta\phi$ correlation function in the 0-20 $\%$ event class 
of p-Pb collisions at $\sqrt{s_{NN} }$ = 5.02 TeV from EPOS 3 considering all particles from core and corona.}
%\end{center}
\label{partondist_XY}
\end{figure}

The ridge is also present in the corona-corona correlation in both 60-100\% and 0-20\% event class as shown in Fig 2(d). In the lowest multiplicity class this may be dominantly due to the flux tube initial contidions. Flux tubes are extended upto many units of space time rapidity as they extend from projectile to target remnants \cite {epos_PbPb_lambdakshort_enhance} and therefore may contribute to the ridge like structure. The contribution from the fluid-jet interaction is expected to be more prominent in the highest multiplicity class as more jet hadrons carry the fluid properties and the ridge is found to be slightly enhanced compared to the lowest multiplicity class as shown in Fig 2(d). 

%The ridge originating from different origins in the 0-20\% multiplicity class are compared in Fig 3. The %dominant contribution is from core-core or pure hydrodynamical origin but contribution from fluid-jet %interaction is non zero.

\begin{figure}[htb!]
%\begin{center}
\includegraphics[scale=0.40]{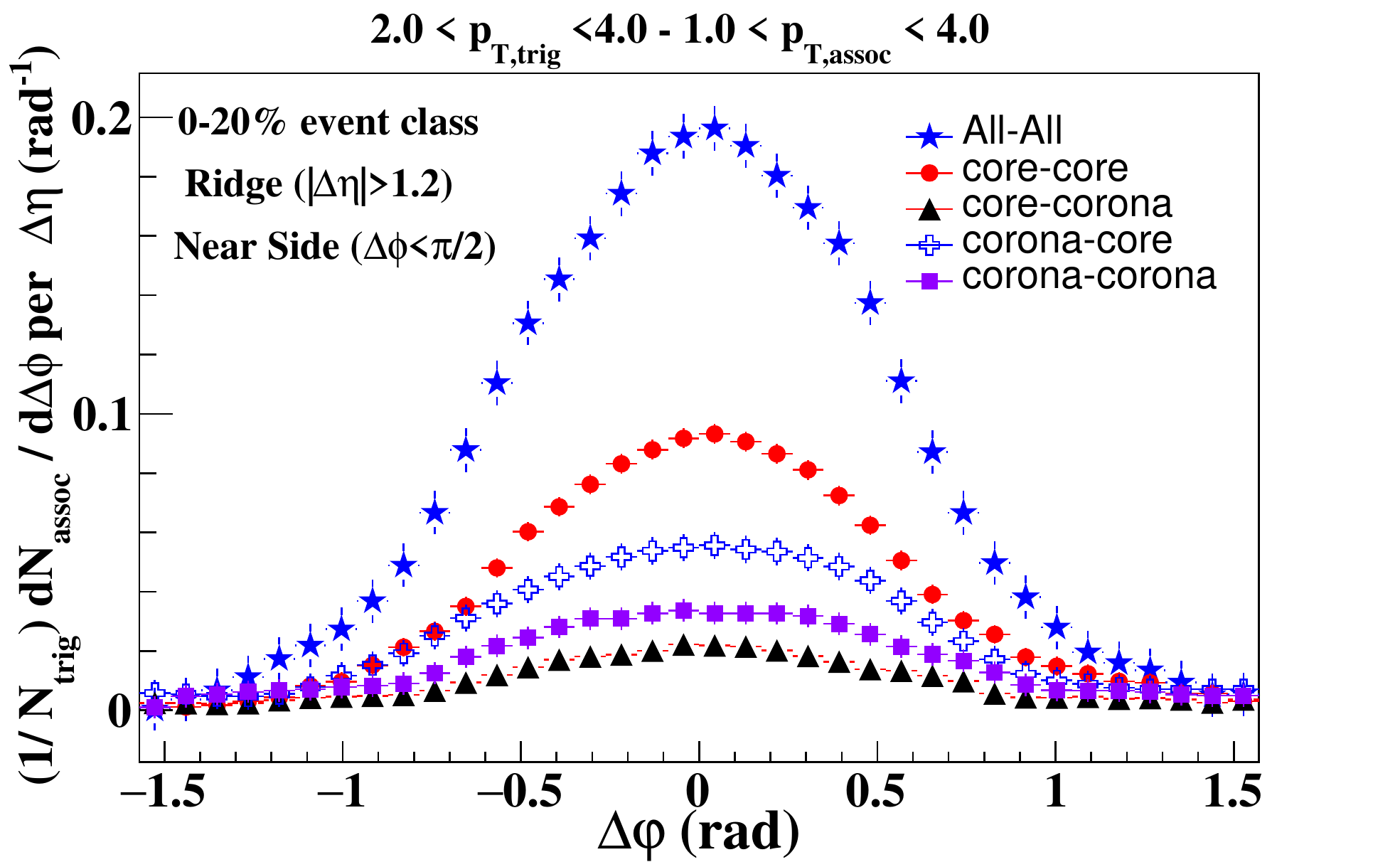}

\caption{[Color online] Relative contribution of ridges from different origins towards the total ridge in the 0-20\% event class of p-Pb collisions at $\sqrt{s_{NN} }$ = 5.02 TeV.}
%\end{center}
\label{partondist_XY}
\end{figure}

\begin{figure}[htb!]
%\begin{center}
\includegraphics[scale=0.42]{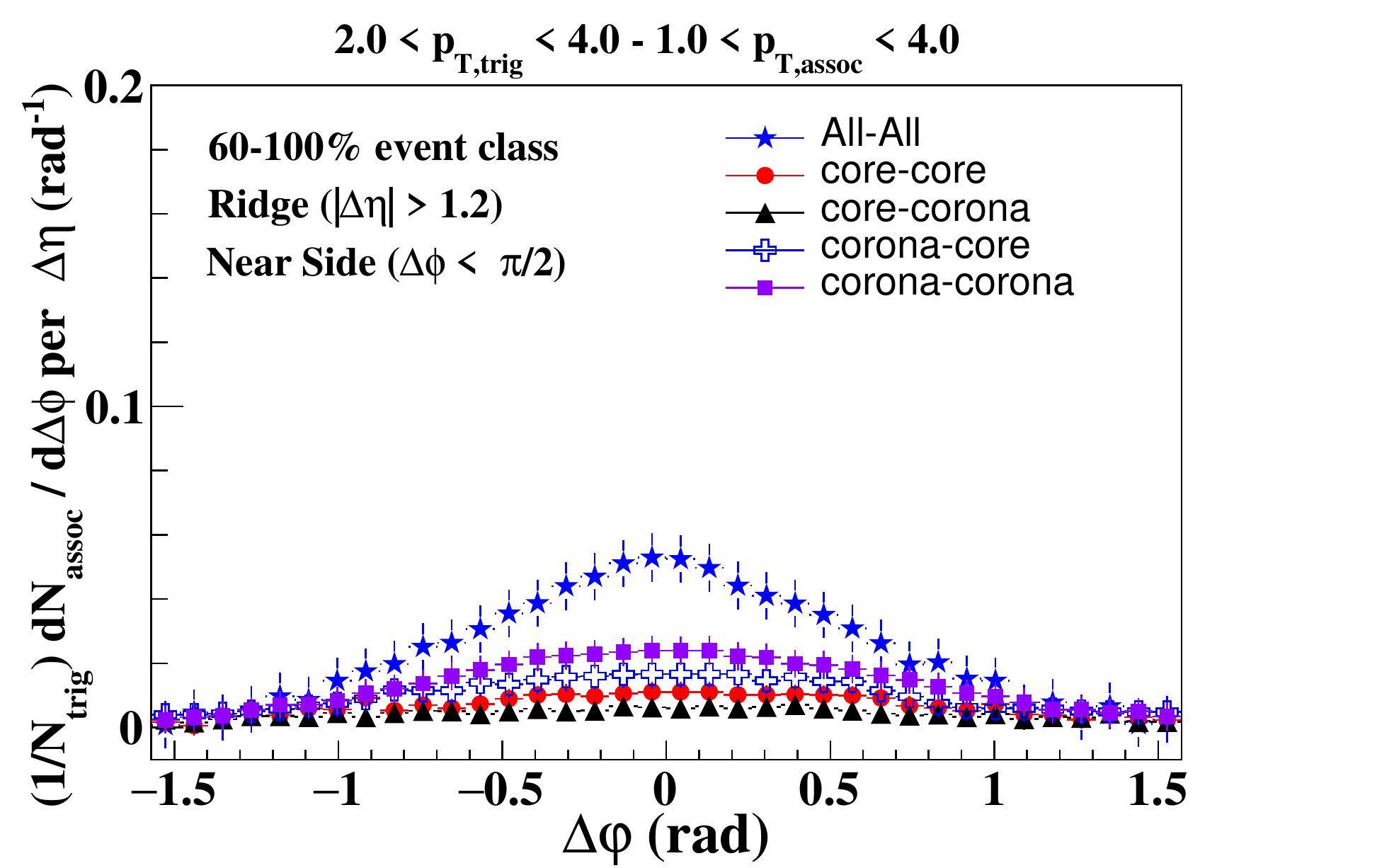}

\caption{[Color online] Relative contribution of ridges from different origins towards the total ridge in the 60-100\% event class of p-Pb collisions at $\sqrt{s_{NN} }$ = 5.02 TeV.}
%\end{center}
\label{partondist_XY}
\end{figure}
%[width=8cm,height=7cm

The 2D correlation structure obtained from the correlation analysis considering all particles from core and corona in the 0-20\% event class is shown in Fig 3 and the ridge structure extracted from this correlation function is shown in Fig 4. It is mentioned as "All-All" in both cases. The relative contribution of the ridges from different origins towards the total ("All-All") ridge in the 0-20\% multiplicity class is shown in Fig. 4 and it is obtained by normalizing the core and corona triggered correlations (shown in Fig 2) by total (core+corona) number of trigger particles in the trigger $p_{T}$ region. In the 0-20\% event class, there are around 68\% particles originating from corona and the rest 32\% are from core compared to the 77\% corona and 23\% core in the 60-100\% event class in the trigger $p_{T}$ region (2.0  $<p{_T}<$ 4.0 GeV/$\it{c}$). So, due to normalization of the core and corona triggered correlations (shown in Fig 2) by all (corona+core) triggers, the core triggered correlations are diluted by 68\% whereas the corona triggered correlations by 32\% in the 0-20\% event class. The relative contribution of core and corona (core : corona) in the associated $p_{T}$ region (1.0  $<p{_T}<$ 4.0 GeV/$\it{c}$) changes from (37\% : 63\%) in 60-100\% event class to (56\% : 44\%) in 0-20\% event class. From Fig 4 it is clear that though the hydrodynamical evolution of the system (core-core correlations) contributes maximum to the total ("All-All") ridge, a non zero contribution from the fluid-jet interaction is also present in the high multiplicity class of p-Pb collisions. In Fig 5, it has been shown that in 60-100\% event class, the total ridge ("All-All") is much smaller than the same in 0-20\% event class and it is mainly dominated by the corona triggered correlations (corona-corona and corona-core) compared to the core triggered ones (core-core and core-corona). This indicates that the effect of core towards ridge structure reduces with decrease in multiplicity and the corona triggered correlations play an important role in the lower multiplicity classes of p-Pb collisions.

Our work shows that in the context of EPOS 3 - which takes into account hydrodynamically expanding bulk matter, jets and the interaction between the two, the ridge has nonzero contribution from the fluid-jet interaction.  As the probability of producing jet hadrons inside or at the surface of the high density of core increases with the multiplicity/system size \cite {epos_PbPb_lambdakshort_enhance}, the ridge contribution from fluid-jet interaction increases with increase in multiplicity as shown in Fig 2. The jet hadrons carrying fluid information play an important role as they are correlated with the bulk part of the system and contributing to the ridge structure. Futher study with identified particles may be helpful to understand the chemical composition of the ridge in the p-Pb collisions at LHC energy.

\section*{Acknowledgements} 
 We acknowledge fruitful discussions and suggestions from Dr. Klaus Werner at various stages of this work. Thanks to VECC grid computing team for their 
constant effort to keep the facility running and helping in EPOS data generation.

%\linenumbers
\end{document}